\documentstyle[preprint,aps,epsfig]{revtex}
\draft
\begin{document}
\title{Excitation function of nucleon and pion elliptic flow in
relativistic heavy-ion collisions}
\bigskip
\author{Bao-An Li\footnote{Email: Bali@navajo.astate.edu}$^{\rm a}$,
C.M. Ko\footnote{Email: Ko@comp.tamu.edu}$^{\rm b}$,
Andrew T. Sustich\footnote{Email: Sustich@navajo.astate.edu}$^{\rm a}$
and Bin Zhang\footnote{Email: Bzhang@kogroup.tamu.edu}$^{\rm b}$}
\address{$^{\rm a}$Department of Chemistry and Physics\\
Arkansas State University, P.O. Box 419\\
State University, AR 72467-0419, USA}
\address{$^{\rm b}$Cyclotron Institute and Physics Department,\\
Texas A\&M University, College Station, TX 77843, USA}

\maketitle

\begin{quote}
Within a relativistic transport (ART) model for heavy-ion collisions,
we show that the recently observed characteristic change 
from out-of-plane to in-plane elliptic flow of protons in mid-central Au+Au 
collisions as the incident energy increases is consistent with the 
calculated results using a stiff nuclear equation of state ($K$= 380 MeV). 
We have also studied the elliptical flow of pions and the transverse
momentum dependence of both the nucleon and pion elliptic flow
in order to gain further insight about the collision dynamics. 
\end{quote}
%\narrowtext
\newpage
The elliptic flow of hadrons in relativistic heavy ion 
collisions has been a subject of great interest as it may reveal 
the signatures of possible QGP phase transition in these collisions 
[see Ref. \cite{oll98} for a recent review]. Based on kinematical and 
geometrical considerations of relativistic heavy-ion collisions, 
Ollitrault \cite{oll92} predicted that as the incident energy increases
nucleons would change from an out-of-plane elliptical flow to an in-plane 
one. Such a transition has recently been observed in collisions 
of heavy ions from the Alternating Gradient Synchrotron (AGS) at 
the Brookhaven National Laboratory \cite{e877,liu98,pin99}. 
Data from the EOS, E895 and E877 collaborations on the proton elliptic
flow in mid-central Au+Au collisions show that the beam energy 
($E_{\rm tr}$) at which the elliptical flow changes
sign is about 4 GeV/A \cite{liu98,pin99}. Studies based on transport 
models have indicated that the value for $E_{\rm tr}$ depends on 
the nuclear equation of state (EOS) at high densities \cite{sorge97,pawel98}. 
Using a relativistic Boltzmann-Equation model (BEM), it has been found
that the experimental data can be understood if the
nuclear equation of state used in the model is stiff ($K$=380 MeV) 
for beam energies below $E_{\rm tr}$ but soft ($K$=210 MeV) 
for beam energies above $E_{\rm tr}$ \cite{pin99}. Since the baryon
density reached in heavy ion collisions at these energies increases 
with the beam energy, the above study thus suggests that the nuclear equation 
of state is softened at high densities.  Such a softened equation of
state may imply the onset of a phase change as suggested by 
lattice studies of the QCD at finite temperature and zero baryon chemical 
potential. However, to put this conclusion on a firm ground requires
further studies using other models. In this Rapid Communication, we 
shall study the elliptical flow in heavy ion collisions at AGS 
energies using a Relativistic Transport (ART) Model \cite{art} and 
show that the experimental data is consistent instead with the 
prediction using a stiff EOS without invoking a softening at high 
densities \cite{atlanta}. Furthermore, we shall show that by studying 
both the nucleon and pion elliptic flow as a 
function of beam energy and transverse momentum one can obtain
much more information about the reaction dynamics and the origin 
of the transition in the sign of elliptic flow.
       
Our study is based on the relativistic transport model ART for
heavy ion collisions. We refer the reader to Ref. \cite{art} for 
details of the model and its applications in studying various 
aspects of relativistic heavy-ion collisions from Bevalac to AGS 
energies. The elliptic flow reflects the anisotropy in the 
particle transverse momentum ($p_t$) distribution at midrapidity, i.e., 
$v_2\equiv <(p_x^2-p_y^2)/p_t^2> $, where the average is taken
over all particles of a given kind in all events \cite{vol98}. 
In the upper window of Fig. 1, we compare the excitation function of 
$v_2$ for protons in mid-central Au+Au reactions obtained 
using the stiff (cross), soft (filled square) EOS and the cascade 
(open square) with the experimental data (open circles) 
of Ref. \cite{pin99}. An impact parameter of 5 fm, which
is consistent with that in the data analysis \cite{pin99,roy98},
is used in the calculations. In agreement with other 
model calculations \cite{sorge97,pawel98,urqmd}, our calculated results 
also show that the transition energy in the
proton elliptic flow is very sensitive 
to the nuclear EOS. The value of $E_{\rm tr}$ 
is more than 4 GeV/A in the case
of a stiff EOS but decreases to below 3 GeV/A for a soft EOS.
As discussed in Ref. \cite{pawel98}, a soft EOS, which gives a smaller 
sound velocity than that of a stiff EOS, reduces the squeeze-out contribution 
and thus leads to a smaller transition energy in proton elliptic flow. 
In the case of cascade calculations, the absence of a repulsive potential 
further reduces the squeeze-out contribution and results in an 
essentially in-plane flow in the beam energy range considered
here.  On the other hand, the value of $v_2$ in our calculations
is insensitive to the nuclear EOS for incident energies above about 
6 GeV/A. This is different from the results of Ref. \cite{pawel98}, where 
a distinct difference is seen between the elliptic flow due to a soft 
and a stiff EOS. Our results also differ from that of Ref. \cite{urqmd}
based on the Ultrarelativistic Quantum Molecular Dynamics (UrQMD),
in which the elliptical flow in the case of a stiff EOS is much smaller
than that from the cascade model even for incident energies above 6
GeV/A. However, in both our study and that from the UrQMD 
the experimental, data are found to be consistent with 
the calculated results using the stiff EOS in this beam energy range.
These results are thus different from that of
Ref. \cite{pin99}, where calculations based on the BEM model
show that the experimental data suggest a softening of the EOS from 
a stiff one at low beam energies to a softer one at higher energies.
Since different model calculations lead to different 
dependence of the proton elliptical flow on the nuclear EOS, it is
thus not possible at present to draw conclusions from comparisons of 
the theoretical results with the experimental data. 
To test these theoretical models, simultaneous studies of 
other experimental observables will be useful.

Since pions are abundantly produced in high energy heavy ion collisions,
their elliptical flow is expected to provide further insight about the
collision dynamics. In the lower window of Fig. 1, we show
our predictions for the excitation function of the pion elliptic flow.
All three charge states of the pion are included in the analysis. 
Effects due to the different charges will be discussed in the next 
paragraph. It is seen that pions also show a transition from out-of-plane to 
in-plane elliptic flow as the beam energy increases. However, 
both the magnitude of pion elliptic flow and the transition 
energy at which it changes sign are significantly smaller than those 
for nucleons. This can be qualitatively understood from the collision 
dynamics. For nucleons, the sign and magnitude of elliptic flow 
depends on both transverse expansion time of participant nuclear matter and 
the passage time of the two colliding nuclei. 
The latter reflects the time scale for the spectators to be effective 
in preventing the participant hadrons from developing an in-plane 
flow, thus enhancing the squeeze-out contribution to the
elliptical flow. For pions, however, the shadowing 
effect due to spectator nucleons is less important
as a result of the time delay in their production, i.e.,
a significant number of pions are emitted later in the reaction 
from the decay of both baryon and meson resonances after the spectator 
nucleons have already moved away. Therefore, both the magnitude of 
squeeze-out contribution and the transition energy in the
pion elliptic flow are significantly smaller than those of nucleons. 
The study of the excitation function of both nucleon and pion flow 
is useful in understanding the origin of the transition 
from out-of-plane to in-plane elliptic flow.

In Fig. 2,  we show the $p_t$ dependence of nucleon and pion elliptic 
flow in a mid-central collision of Au+Au
at a beam momentum of 6 GeV/c. They are obtained from the ART model 
with a soft nuclear EOS. For protons, the elliptic flow
increases approximately quardratically at low $p_t$ and then increases
linearly at high $p_t$, as expected from the
nucleon azimuthal angle distributions \cite{oll98,pawel95,hei99}. 
For pions, their $v_2$ value is larger than that for protons at low $p_t$ 
but becomes similar at high $p_t$.
Again, one can understand this result from the reaction dynamics.
Low $p_t$ pions are more likely produced later in the reaction, 
and they are thus less likely to be shadowed by 
spectator nucleons and have thus a larger in-plane flow 
compared to low $p_t$ protons. On the other hand, high
$p_t$ pions are mainly produced early in the reaction
and thus freeze out together with high $p_t$ nucleons, leading then 
to a similar elliptic flow, which approaches that of the hydrodynamical 
limit \cite{hei99}. It is interesting 
to mention that the observed $p_t$ dependence of $v_2$ 
for nucleons and pions is remarkably similar to what one 
found at both Bevalac/SIS \cite{pawel95,li93} and 
SPS energies \cite{na49}, indicating the similarity of the 
collision dynamics at these different energies.
We note that negative pions have higher
in-plane flow than the positive ones as a result of the 
Coulomb potential from protons, i.e., 
negative pions are attracted to while positive ones are repelled 
away from the reaction plane by protons. 

In summary, using a relativistic transport model we have found that
the transition from out-of-plane to in-plane elliptic flow
in mid-central Au+Au collisions as the beam energy increases
is consistent with a stiff nuclear EOS without invoking a phase transition. 
This result is consistent with that from the UrQMD model
but different from that from the BEM model. To help disentangle these
different predictions, we have also showed the excitation function of the
pion elliptic flow and the transverse momentum dependence of both 
the nucleon and pion elliptic flow, which are expected to 
reveal interesting information about both the reaction dynamics and 
the origin of the observed change in the sign of elliptic flow. 
  
This work was supported in part by NSF Grant No. PHY-9870038, the
Robert A Welch foundation under Grant A-1358, and the Texas
Advanced Research Program.

\newpage

\begin{figure}[h]
\vspace{1in}
\centerline{\epsfig{file=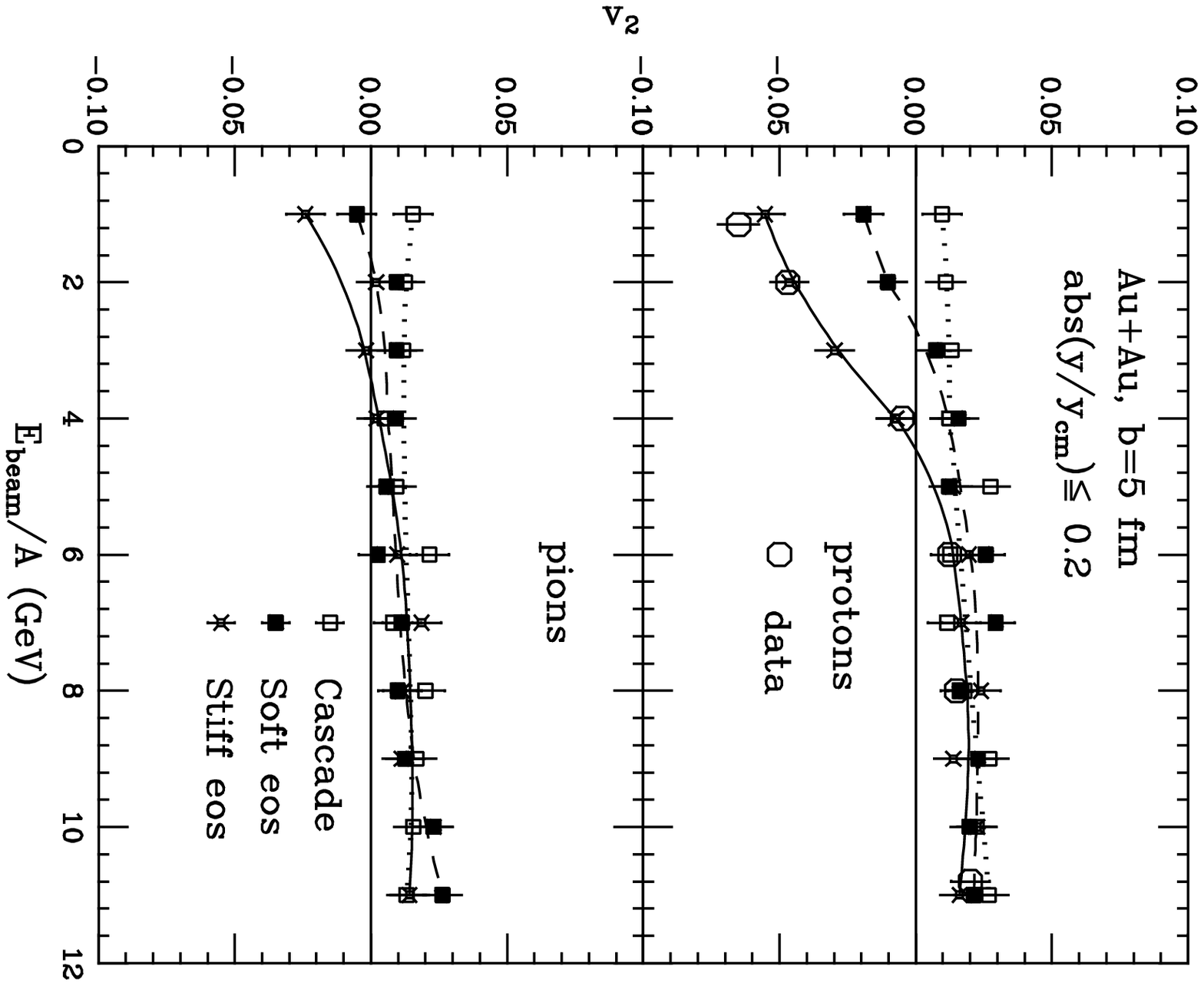,width=4in,height=4in,angle=90}}
\vspace{0.3in}
\caption{The excitation function of proton (upper) and pion (lower)
elliptic flow in mid-central Au+Au reactions.}
\label{fig1} 
\end{figure}

\newpage

\vspace{12in}
\begin{figure}[h]
\centerline{}
\vspace{1in}
\centerline{\epsfig{file=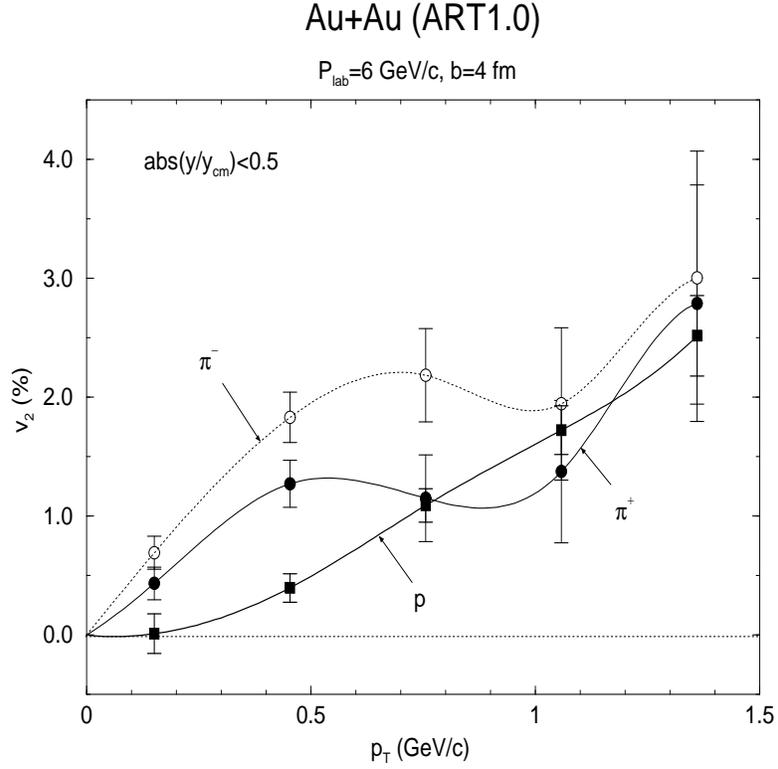,width=4in,height=4in,angle=-90}}
\vspace{0.3in}
\caption{The transverse momentum dependence of nucleon and pion 
elliptic flow in the reaction of Au+Au at $p_{\rm beam}/A=6$ GeV/c
and an impact parameter of 4 fm using a soft nuclear equation of
state.}
\label{fig2} 
\end{figure}

\end{document}